\documentstyle[aps,prl,twocolumn]{revtex}
\input epsf

\begin{document}
\draft
\title{Direct observation of irrotational flow and evidence of superfluidity in a rotating Bose-Einstein condensate}
\author{G. Hechenblaikner,  E. Hodby, S.A. Hopkins, O.M. Marag\`o, and C.J. Foot}
\address{Clarendon Laboratory, Department of Physics, University of Oxford,\\
Parks Road, Oxford, OX1 3PU, \\
United Kingdom.}
\date{\today}

\maketitle

\begin{abstract}
We observed the expansion of vortex-free Bose-condensates after
their sudden release from a slowly rotating anisotropic trap. Our
results show clear experimental evidence of the irrotational flow
expected for a superfluid. The expansion from a rotating trap has
strong features associated with the superfluid nature of a
Bose-condensate, namely that the condensate cannot at any point be
cylindrically symmetric with respect to the axis of rotation since
such a wavefunction cannot possess angular momentum. Consequently,
an initially rotating condensate expands in a distinctively
different way to one released from a static trap. We report
measurements of this phenomenon in absorption images of the
condensate taken along the direction of the rotation axis.
\end{abstract}

In their recent theoretical paper, Edwards et al. \cite{edwards}
calculated the time evolution of the expansion of a slowly
rotating BEC. At higher rotation rates, greater than about half
the radial trap frequency, the superfluid nature of the condensate
has been demonstrated strikingly in the formation of quantised
vortices \cite{madison}. However they point out that the
superfluidity of the condensate, together with the conservation of
angular moment, also leads to clearly observable consequences at
lower rotation rates, under conditions where no vortices are
present. We report here the experimental verification of these
predictions by time-of-flight measurements on a condensate
released from a rotating anisotropic trap \cite{arlt}. These data
give further evidence for the superfluidity of Bose-condensed
gases. Other recent experiments on this important aspect of
Bose-condensation include the observation of quantized vortices
\cite{matthews,madison}, the absence of dissipation below a
critical velocity \cite{raman} and the quenching of the moment of
inertia in the scissors mode \cite{marago}.

The solutions of the hydrodynamic equations for superfluids (that
also describe the scissors mode \cite{stringari}) show that a
condensate always has a moment inertia less than that of a rigid
body of the same mass distribution \cite{zambelli}. For the
special case of cylindrical symmetry, the superfluid has zero
moment of inertia about the symmetry axis. Thus a condensate with
some angular momentum can never have a circular shape when viewed
along the rotation axis, since this would imply the unphysical
situation of infinite angular velocity. An initially rotating
condensate therefore expands in a different way to one released
from static trap. We observed this phenomenon in absorption images
of the condensate taken along the direction of the rotational axis
-  the projection of the cloud in this direction never becomes
circular (an aspect ratio of unity). The evolution of the
condensate density distribution is calculated using the same
hydrodynamic equations as in \cite{edwards}, and agrees well with
our data. We give a brief review of the underlying theoretical
aspects before describing the experimental procedure and results.

\begin{figure}
\begin{center}\mbox{ \epsfxsize 3.2 in \epsfbox{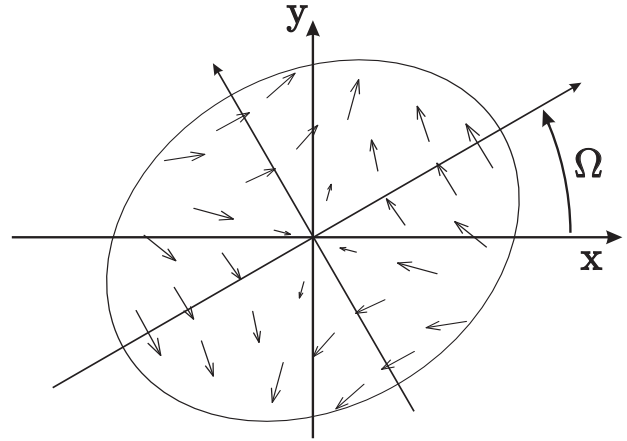}}\end{center}
\caption{The velocity field of a vortex-free condensate in an
elliptical trap, rotating at frequency $\Omega$. The small arrows
of the velocity field indicate the direction of the quadrupolar
flow.} \label{figure1}
\end{figure}

In our experiment we have an anisotropic harmonic potential with
three angular frequencies $\omega_x<\omega_y<\omega_z$. The
potential rotates about the z-axis with angular frequency
$\Omega$. It is convenient to define two new parameters: the trap
deformation,
$\lambda=(\omega_x^2-\omega_y^2)/(\omega_x^2+\omega_y^2)$ and the
mean of the frequencies in the plane of rotation defined as
$\omega_m^2=(\omega_x^2+\omega_y^2)/2$. In the hydrodynamic limit,
a condensate rotating in this trap displays a quadrupolar flow
pattern, as described in fig.\ref{figure1}, and has a wavefunction
which looks like:
\begin{equation}
\Psi({\bf r})=\sqrt{\rho({\bf r})} e^{i\frac{m\nu}{\hbar}xy},
\label{wavefunction}
\end{equation}
where $\rho({\bf r})$ is the condensate density given in
Eq.\ref{density} below. The quadrupole frequency, $\nu$, can be
found from the solutions of the following cubic
equation\cite{anglin}:

\begin{equation}
\tilde{\nu}^3+(1-2\tilde{\Omega}^2)\tilde{\nu}+\lambda\tilde{\Omega}=0,
\label{cubicequation}
\end{equation}
where we introduced the dimensionless quantities
$\tilde{\nu}=\nu/\omega_m$ and $\tilde{\Omega}=\Omega/\omega_m$.

In a rotating trap, the condensate experiences effectively
different trapping frequencies due to its quadrupolar motion and
the condensate density can be written as:

\begin{equation}
\rho({\bf r})=\frac{\mu}{g}\left[1-\frac{m}{2}(\tilde{\omega}_x^2
x^2+\tilde{\omega}_y^2 y^2+\omega_z^2 z^2)\right]. \label{density}
\end{equation}
The modified frequencies $\tilde{\omega}_x$ and $\tilde{\omega}_y$
are given by:

\begin{eqnarray}
\tilde{\omega}_x^2 &=& \omega_x^2+\nu^2-2\nu\Omega\nonumber\\
\tilde{\omega}_y^2 &=& \omega_y^2-\nu^2-2\nu\Omega
\label{effectivefreqs}
\end{eqnarray}
where $\nu$ is found from Eq.\ref{cubicequation}.

Hence, the aspect ratio of the condensate in the trap is
\begin{equation}
\frac{R_x}{R_y}=\frac{\tilde{\omega}_y}{\tilde{\omega}_x}=\sqrt{\frac{\Omega+\nu}{\Omega-\nu}},
\end{equation}
where $R_x,R_y$ are the sizes of the condensate in the x and
y-direction, respectively. The effective chemical potential and
condensate sizes in the trap can also be calculated from
Eqs.\ref{effectivefreqs}.

To calculate the expansion of a condensate when released from the
anisotropic harmonic potential,  we use the following ansatz of
spheroidal form for the condensate density $\rho({\bf r},t)$ and
the velocity field, ${\bf v}({\bf r},t)$, (as in \cite{edwards}):
\begin{eqnarray}
\rho({\bf r},t) &=&
\rho_0(t)-\rho_x(t)x^2-\rho_y(t)y^2-\rho_z(t)z^2-\rho_{xy}(t)
xy\label{densityquadratic}\\ {\bf v}({\bf
r},t)&=&\frac{1}{2}\nabla\left(
v_x(t)x^2+v_y(t)y^2+v_z(t)z^2+v_{xy}(t)xy\right)
\end{eqnarray}
Inserting this ansatz into the hydrodynamic equations of
superfluids yields a set of nine coupled differential equations
for the expansion parameters, which we integrate numerically. The
initial conditions for the density are found from
Eq.\ref{density}. Note that the cross-term $\rho_{xy}(0)$ is
assumed to be zero at the instant when the condensate is released
from the trap. During the evolution in time of flight, this term
grows and can become comparable in size to $\rho_{x}$, so that the
elliptical condensate rotates with respect to its initial release
angle. All components of the velocity field are initially zero,
except for the cross term, which is the quadrupole velocity field
of the rotating condensate, given by $v_{xy}=2\nu$ (see
Eq.\ref{wavefunction}).
 The behavior of the condensate is thus completely
determined by the above initial conditions and the nine coupled
differential equations \cite{edwards}.

The angle of the condensate and its sizes can be found by
diagonalising the quadratic form defined in
Eq.\ref{densityquadratic}. One finds, in time of flight, that at
first the condensate expands along its smaller axes until the
aspect ratio has reached a critical value close to spherical. The
condensate has a quadrupolar flow pattern and an associated
angular momentum. Conservation of this angular momentum does not
allow the condensate to become spherical, as a spherical
quadrupolar flow pattern has no angular momentum. Thus, when
approaching this critical value in the aspect ratio, the
condensate rotates quickly and continues to expand along its
original long axis. The rapid increase in angular velocity close
to the critical aspect ratio arises from the decrease of the
moment of inertia when the aspect ratio approaches spherical
\cite{edwards}. However, energy conservation limits how fast the
condensate can rotate and thus the aspect ratio cannot become
smaller than a certain value. This behaviour is in contrast to
that of a thermal gas which can become spherical and preserve
angular momentum, since there are no constraints of
irrotationality. Thus it is essentially the superfluid nature of
the Bose-condensate which prevents it from becoming spherical.

\begin{figure}
\begin{center}\mbox{ \epsfxsize 2 in \epsfbox{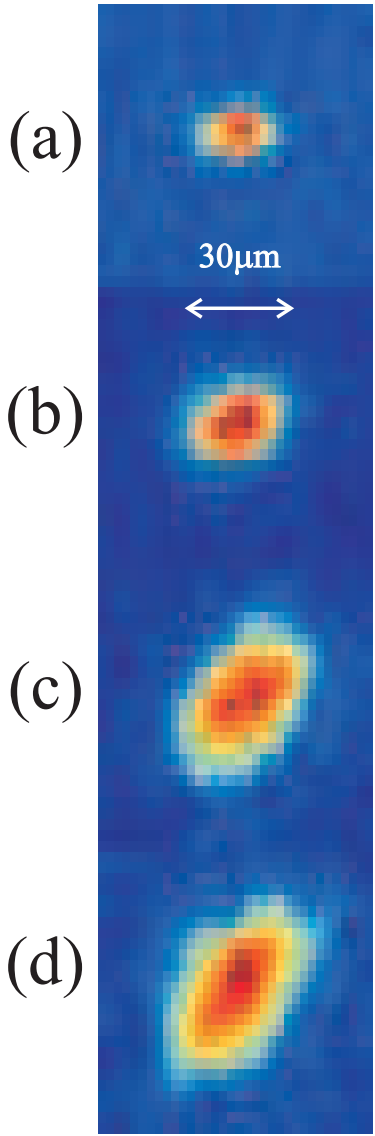}}\end{center}
\caption{Typical images of the condensate at different times after
release from a trap rotating at 28Hz - (a) $1.09$ ms (b) $3.97$ ms
(c) $12.00$ ms (d) $16.06$ ms. At the instant of release the long
axis of the cloud was horizontal. } \label{figure2}
\end{figure}

To observe this behaviour we evaporatively cooled $^{87}$Rb atoms
in a TOP trap to a temperature of $0.5 \, T/T_c$, where $T_c$ is
the critical temperature for condensation. This produced
Bose-condensates of $1.5\times 10^4$ atoms, in a trapping
potential with frequencies $\omega_x/2\pi=\omega_y/2\pi=124$ Hz
and $\omega_z/2\pi=350$ Hz. We then made the trap elliptical with
$\omega_y/\omega_x=1.4$ (corresponding to $\lambda=0.32$) by
changing the ratio of the two TOP-field components to
$B_x/B_y=4.2$. The eccentric trap was rotated by modulating the
high frequency sinusoidal TOP signal ($7$ kHz frequency) at the
low trap rotation frequency $\Omega$, as described in \cite{arlt}.
The condensate was spun up in $500$ ms during which time the
eccentricity was ramped up from zero to its final value. Then the
condensate was left in the rotating trap for another $500$ ms
before it was released at a well defined point in the trap
rotation, from which the angles of the expanding cloud were
measured. Figure \ref{figure2} shows typical absorption images of
the condensate, taken along the axis of rotation, after different
expansion times. For these pictures the trap was rotating at 28 Hz
and at the instant of release the long axis of the cloud was
horizontal. The angle and aspect ratio of the cloud was obtained
from a 2D parabolic fit to the density distribution. Figure
\ref{figure2}b shows the minimum aspect ratio of 1.31, reached
after $\sim 4$ ms of expansion. The angle increases steadily to an
asymptotic value of $\sim 55$ degrees in fig.\ref{figure2}d.

\begin{figure}
\begin{center}\mbox{ \epsfxsize 3.2 in \epsfbox{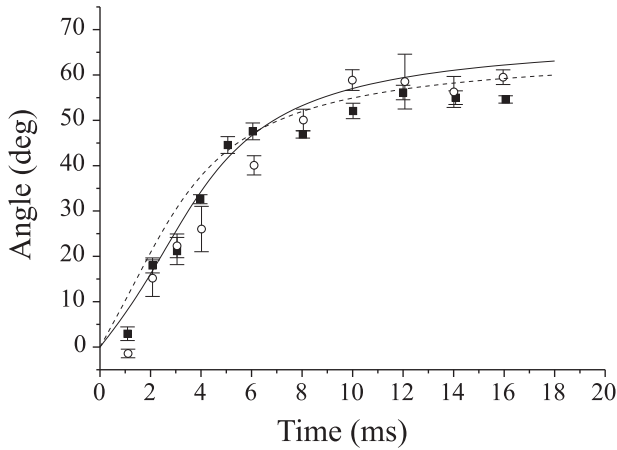}}\end{center}
\caption{The angle of the
condensate plotted against the time of flight. The open circles
and the filled squares denote the measured angles after release
from a trap rotating at $\Omega/2\pi=20$ and $\Omega/2\pi=28$ Hz,
respectively. The solid and the dotted line are the respective
theoretical calculations. } \label{figure3}
\end{figure}

It was necessary to go to a high trap eccentricity so that a clear
deformation of the cloud could be observed. We investigated the
evolution of the condensate in time of flight at two different
trap rotation frequencies $\Omega/2\pi=20$ and $28$ Hz. In both
cases the condensate was released from a trap of $\omega_x /
2\pi=60$ Hz, $\omega_y/2\pi=1.4\times 60$ Hz and
$\omega_z/2\pi=206$ Hz. Fig.\ref{figure3} shows the calculated
change of angle of a condensate in time of flight after being
released from a trap rotating at $\Omega/2 \pi =20$ Hz (solid
line) and $\Omega/2 \pi=28$ Hz (dotted line), with the
experimental data points superimposed. In both cases the angle
evolves in a similar manner and reaches $45$ degrees after about
$6$ ms. After $18$ ms the angle assumes an asymptotic value
between $55$ and $60$ degrees.

\begin{figure}
\begin{center}\mbox{ \epsfxsize 3.2 in \epsfbox{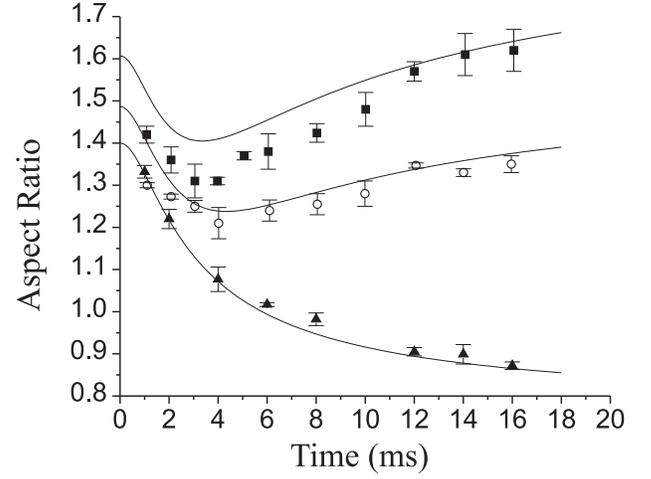}}\end{center}
\caption{The aspect ratio of a condensate in time of flight.
Initially rotating condensates (upper and middle theoretical
curves) exhibit a strong backbending effect after $4$ ms and the
condensate never becomes circular. However, after release from the
non-rotating trap, the aspect ratio decreases steadily and inverts
(lower curve); it is unity at about $6$ ms.
\\ Upper curve and filled squares, $\Omega/2 \pi =$ 28 Hz
\\ Middle curve and open circles, $\Omega/2 \pi =$ 20 Hz
\\ Lower curve and filled triangles, Static trap. } \label{figure4}
\end{figure}

Figure \ref{figure4} shows the theoretical prediction for the
evolution of the condensate's aspect ratio in time of flight, with
experimentally measured values superimposed, for $\Omega/2\pi=28$
Hz (upper curve), for $\Omega/2\pi=20$ Hz (middle curve) and for
release from a nonrotating trap (lower curve).  The data clearly
demonstrate how the aspect ratio for an initially rotating
condensate decreases up to a critical point, which is reached
after approximately $4$ ms. From that point on it does not
continue to expand along its minor axis but the aspect ratio
increases again because the condensate cannot become circular
under these conditions. However, the condensate released from a
static trap has no velocity field which prevents it from becoming
circular and the aspect ratio decreases steadily from
$\omega_y/\omega_x$ to a final value of less than one. Every
experimental point displayed is the average of several
measurements. For each time of flight we had to refocus our
imaging system as the atoms move out of focus under gravity when
released from the trap. However incorrect focusing would only
result in a more circular image and the minimum value for the
measurement of the aspect ratio of rotating condensates was never
consistent with unity. There is remarkable agreement between the
experimental data and the theoretical predictions. However, we
observed a small deviation of the experimental data from the
predicted values for the higher rotation frequency
($\Omega/2\pi=28$ Hz, shown in the upper curve). We do not know
the origin of that deviation.

Our results show that an expanding vortex-free Bose-condensate
with some angular momentum refuses to become circular about the
axis of rotation, as predicted by Edwards et al \cite{edwards}.
This provides direct evidence that Bose-condensed gases flow
irrotationally as a consequence of their superfluidity.  These
measurements complement the observations of the scissors mode
\cite{marago} where the irrotational flow was deduced from
measurements of the oscillation frequencies of a trapped
condensate - a superfluid has less moment of inertia than a
classical object, of the same size, undergoing rigid body motion
hence the superfluid oscillates faster. Other measurements of the
flow of a Bose-condensed gas would be of great interest, for
example flow through a narrow tube \cite{jackson}, or array of
holes, analogous to the superleak experiments with liquid helium.

We thank M. Edwards, S. Stringari and the members of the Oxford
theoretical BEC group for fruitful discussions. We are very
grateful to J. Arlt for his contribution to the early stages of
this experiment. We acknowledge support from the EPSRC, St. John's
College, Oxford (G.H.), Christ Church College, Oxford (E.H.),
Linacre College, Oxford, MCFA and the EC (O.M.).



\end{document}